# Surface solitons at interfaces of arrays with spatially-modulated nonlinearity

Y. V. Kartashov,[1] V. A. Vysloukh,[2] A. Szameit,[3] F. Dreisow,[3] M. Heinrich,[3] S. Nolte,[3] A. Tünnermann,[3] T. Pertsch,[3] and L. Torner[1]

[1]ICFO-Institut de Ciencies Fotoniques, and Universitat Politecnica de Catalunya, Mediterranean Technology Park, 08860 Castelldefels (Barcelona), Spain

[2]Universidad de las Americas – Puebla, 72820, Puebla, Mexico

[3]Institute of Applied Physics, Friedrich-Schiller-University Jena, Max-Wien-Platz 1, 07743 Jena, Germany

We address the properties of two-dimensional surface solitons supported by the interface of a waveguide array whose nonlinearity is periodically modulated. When the nonlinearity strength reaches its minima at the points where the linear refractive index attains its maxima, we found that nonlinear surface waves exist and can be made stable *only within a limited band* of input energy flows, and for lattice depths exceeding a lower threshold.

Guided waves supported by the nonlinear interfaces of different uniform materials have been under active investigation since the 1980s due to their unique physical properties (see Refs. [1-3] for reviews). A number of applications for such guiding structures have been suggested, including the implementation of optical limiters, bistable devices, all-optical couplers and switches [4]. On the other hand, a periodicity of material may substantially affect properties of guided waves [5], including surface waves. Progress in the fabrication of periodic waveguide arrays opened the route to observation of surface waves at reduced power levels. Different types of one- [6-11] and two-dimensional [12-19] surface lattice solitons have been predicted and observed. In most previous studies, nonlinearity was spatially uniform inside the material. Nevertheless, in actual practice the spatial profile of the nonlinearity can be controlled, too. This is the case, e.g., of arrays fabricated by Ti-indiffusion in LiNbO crystals [10,11] where one may fine-tune the nonlinearity profile by changing the concentration of dopants. Another example is given by arrays written in glass by high-intensity fs laser pulses [20,21], where optical damage produced by a tightly focused laser beam results in an increase of the linear refractive index accompanied by a simultaneous decrease of the nonlinear coefficient.

In this Letter we address surface solitons at the edge of a two-dimensional semi-infinite waveguide array whose nonlinearity is periodically modulated in such a way that the nonlinear coefficient takes its minimal value at the points where the linear refractive index reaches its maxima. We show that surface waves in this system can be stable *only inside a limited band* of propagation constants and energy flows.

Our model is based on the nonlinear Schrödinger equation for the amplitude $q$ of a beam propagating along the interface of a two-dimensional optical lattice:

$$i\frac{\partial q}{\partial \xi} = -\frac{1}{2}\left(\frac{\partial^2 q}{\partial \eta^2} + \frac{\partial^2 q}{\partial \zeta^2}\right) - [1 - \sigma R(\eta,\zeta)]|q|^2 q - pR(\eta,\zeta)q. \qquad (1)$$

Here the transverse $\eta, \zeta$ and the longitudinal $\xi$ coordinates are scaled to the beam width and the diffraction length, respectively; the parameters $p$ and $\sigma$ characterize the depths of modulation of the refractive index and nonlinearity, respectively, while the profile of the refractive index is given by $R(\eta,\zeta) = \sum_{m=-\infty}^{\infty}\sum_{k=0}^{\infty} \exp[-(\eta - kw_s)^2/w_\eta^2 - (\zeta - mw_s)^2/w_\zeta^2]$. The values $w_\eta, w_\zeta$ account for the waveguide widths and $w_s$ represents the waveguide spacing. Thus, the nonlinear coefficient $1 - \sigma R$ attains its minima at the points where the refractive index has a maximum. This situation, corresponding to an out-of-phase spatial modulation of refractive index and nonlinearity, appears when the lattice is written by fs laser pulses [22,23]. Here we set $w_\eta = w_\zeta = 1/2$ and $w_s = 2$. Equation (1) conserves the energy flow $U$ and Hamiltonian $H$:

$$\begin{aligned}U &= \int\int_{-\infty}^{\infty} |q|^2\, d\eta d\zeta, \\ H &= \frac{1}{2}\int\int_{-\infty}^{\infty}[|\partial q/\partial \eta|^2 + |\partial q/\partial \zeta|^2 - 2pR|q|^2 - (1-\sigma R)|q|^4]d\eta d\zeta.\end{aligned} \qquad (2)$$

We search for fundamental soliton solutions of Eq. (1) residing in the near-surface channels of semi-infinite lattice in the form $q(\eta,\zeta,\xi) = w(\eta,\zeta)\exp(ib\xi)$, where $b$ is the propagation constant. To analyze the stability of such states we solve the linear eigenvalue problem

$$\delta u = -\frac{1}{2}\left(\frac{\partial^2 v}{\partial \eta^2} + \frac{\partial^2 v}{\partial \zeta^2}\right) + bv - (1-\sigma R)vw^2 - pRv,$$
$$\delta v = \frac{1}{2}\left(\frac{\partial^2 u}{\partial \eta^2} + \frac{\partial^2 u}{\partial \zeta^2}\right) - bu + 3(1-\sigma R)uw^2 + pRu, \quad (3)$$

for the perturbation components $u, v$. Equations (3) were obtained upon substitution of the perturbed light field $q = [w + u\exp(\delta\xi) + iv\exp(\delta\xi)]\exp(ib\xi)$ into Eq. (1) and linearization around the stationary solution $w$. The solution $w$ is stable if the real part $\delta_r$ of the perturbation growth rate $\delta$ vanishes.

Typical profiles of the surface solitons are shown in Fig. 1. As in the case of a usual lattice $(\sigma=0)$, low-amplitude solitons strongly expand into the lattice bulk and only penetrate weakly into the uniform medium [Fig. 1(a)]. Increasing the energy flow results in a localization of light in the near-surface channel [Fig. 1(b)]. Nevertheless, in contrast to media with $\sigma = 0$, a further growth of the field amplitude results in a faster increase of the nonlinear contribution to the refractive index in the space between the rows, where $1 - \sigma R$ has a local maximum. Thus, the structure is characterized by the competition between linear refraction that tends to trap light inside the waveguides and spatially inhomogeneous self-focusing that causes light concentration between the waveguides. Since nonlinear effects dominate when the soliton amplitude becomes sufficiently high, the large-amplitude soliton shifts into the region between the first and second waveguide rows [Fig. 1(c)]. This is an entirely surface effect stemming from the modulation of nonlinearity only in the half-space $\eta > 0$, which give rise to a preferable direction of the soliton shift.

The competition between linear refraction and self-action results in a nontrivial $U(b)$ dependence [Fig. 2(a)]. For $b$ values close to the cutoff $b_{\text{co}}$, one has a non-monotonic dependence $U(b)$, which is typical for surface waves. However, in contrast to lattices with $\sigma = 0$, where far from the cutoff the energy flow $U$ increases with $b$ and asymptotically approaches the value $U = 5.85$, we found that at interfaces with a non-uniform nonlinearity the energy flow decreases when $b$ exceeds a certain critical value. It is inside this region the soliton center gradually shifts into the space between the first and second rows.

In accordance with the Vakhitov-Kolokolov criterion surface solitons are stable only in the region where $dU/db > 0$. Therefore, the stability domain is bounded between a certain minimal $U_{\min}$ and maximal $U_{\max}$ energy flows. The dependence of the Hamiltonian on $U$ exhibits two cuspidal points, accounting for the existence of a single stability domain and two instability domains [Fig. 2(b)]. The stability domains for surface solitons are shown on the planes $(\sigma,U)$ and $(p,U)$ on Figs. 2(c) and 2(d), respectively. The minimal energy flow $U_{\min}$ monotonically increases with $\sigma$, while $U_{\max}$ initially grows, but then remains almost constant. The stability domain vanishes completely when $\sigma$ exceeds the critical value $\sigma_{\rm cr}$. This occurs usually for $\sigma > 1$, which physically correspond to a transition from the focusing to a defocusing nonlinearity in the very centers of the waveguides and can hardly be achieved in practice. The stability domain expands with an increase of the lattice depth and it vanishes completely when $p$ is below the critical value $p_{\rm cr}$ [Fig. 2(d)]. While solitons still exist for $\sigma > \sigma_{\rm cr}$ and $p < p_{\rm cr}$, they are shifted into the space between first and second waveguide rows and are always unstable. The cutoff $b_{\rm co}$ increases with $p$ in lattices which are deep enough [Fig. 2(e)]. We also tested the stability numerically, by solving the Eqs. (3), and found that all unstable branches are associated with exponential instabilities [see Fig. 2(f) for $\delta_r(b)$ dependence].

The spatial modulation of the nonlinearity has a remarkable impact on the process of surface waves excitation. To understand the specific features of this process, it is instructive to consider the shift of the soliton center along the $\eta$-axis with increase of $U$ [Fig. 3(a)]. For both, low and high $U$, the soliton center shifts into the lattice depth, and only for intermediate values of $U$ it approaches the interface. Upon dynamical excitation the center of the input Gaussian beam $A\exp(-\eta^2 - \zeta^2)$ follows similar trends. If the input energy flow is too low the beam diffracts rapidly and almost all light goes into the lattice depth [see Fig. 3(b), where we plot the output energy flow passing through the circular aperture of radius $w_{\rm s}/2$ centered at the axis of the launching channel at $\xi = 16$ as a function of $U$]. At intermediate $U$ one achieves an effective excitation of surface soliton, when almost all input energy remains trapped in the vicinity of the launching channel. If the energy flow is too high the input beam drifts into the space between the first and second lattice rows, where it collapses. The corresponding dependence $U_{\rm out}(U)$ appears to be very sharp, while minimal and maximal input energy flows are very close to those for stationary solitons. This suggests a possibility of engineering an all-optical limiter incorporating the interface of the lattice with spatially modulated nonlinearity.

Summarizing, we presented new properties of fundamental surface solitons arising at the interface of optical lattices with an out-of-phase spatial modulation of refractive index and nonlinearity. The competition between linear refraction and self-focusing in the inhomogeneous nonlinearity landscape results in the appearance of restrictions on both minimal and maximal energy flows of stable surface solitons in such a system.

# References with titles

# References without titles

# Figure captions

Figure 1. Profiles of surface solitons with (a) $b = 0.8$, (b) $b = 1.7$, and (c) $b = 3.2$ at $p = 3$, $\sigma = 0.6$. The thick red line indicates the interface position. The blue line in (c) indicates soliton center shift relative to the interface.

Figure 2. (a) Energy flow versus propagation constant and (b) Hamiltonian versus energy flow at $p = 3$, $\sigma = 0.6$. The points marked by circles in (a) correspond to profiles as shown in Fig. 1. Black lines show stable branches, while red lines correspond to unstable branches. Stability domains (c) on the plane $(\sigma,U)$ at $p = 3$ and (d) on the plane $(p,U)$ at $\sigma = 0.7$. (e) The cutoff versus lattice depth at $\sigma = 0.7$. (f) The real part of the perturbation growth rate versus propagation constant at $p = 3$, $\sigma = 0.6$.

Figure 3. (a) Position of the integral soliton center along the $\eta$ axis versus energy flow. Red lines show the unstable branches, black lines show the stable branches, the arrow shows the direction in which the propagation constant increases. (b) The output energy flow concentrated within a ring of radius $w_\mathrm{s}/2$ at $\xi = 16$ versus input energy flow upon surface wave excitation by a Gaussian beam $A\exp(-\eta^2 - \zeta^2)$. Red circles indicate minimal and maximal input energy flows at which effective surface wave excitation occurs. In all cases $p = 3$ and $\sigma = 0.6$.

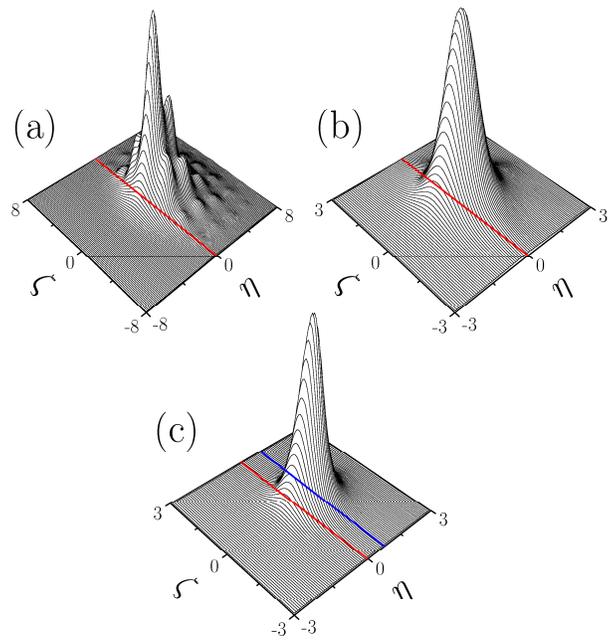

Figure 1.   Profiles of surface solitons with (a) $b = 0.8$, (b) $b = 1.7$, and (c) $b = 3.2$ at $p = 3$, $\sigma = 0.6$. The thick red line indicates the interface position. The blue line in (c) indicates soliton center shift relative to the interface.

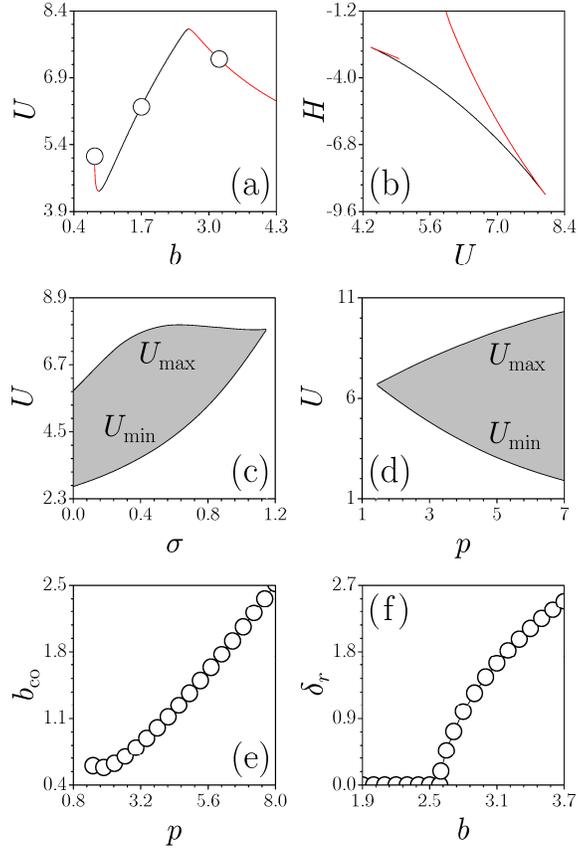

Figure 2. (a) Energy flow versus propagation constant and (b) Hamiltonian versus energy flow at $p=3$, $\sigma=0.6$. The points marked by circles in (a) correspond to profiles as shown in Fig. 1. Black lines show stable branches, while red lines correspond to unstable branches. Stability domains (c) on the plane $(\sigma,U)$ at $p=3$ and (d) on the plane $(p,U)$ at $\sigma=0.7$. (e) The cutoff versus lattice depth at $\sigma=0.7$. (f) The real part of the perturbation growth rate versus propagation constant at $p=3$, $\sigma=0.6$.

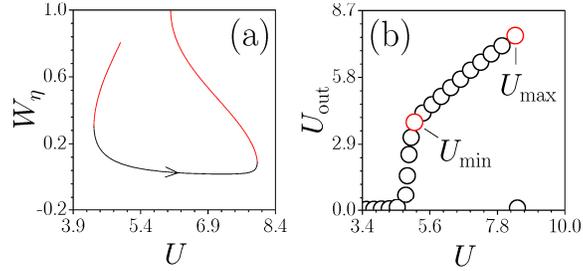

Figure 3.  (a) Position of the integral soliton center along the $\eta$ axis versus energy flow. Red lines show the unstable branches, black lines show the stable branches, the arrow shows the direction in which the propagation constant increases. (b) The output energy flow concentrated within a ring of radius $w_{\rm s}/2$ at $\xi=16$ versus input energy flow upon surface wave excitation by a Gaussian beam $A\exp(-\eta^2-\zeta^2)$. Red circles indicate minimal and maximal input energy flows at which effective surface wave excitation occurs. In all cases $p=3$ and $\sigma=0.6$.